# ANALYSIS OF NEIGHBOUR AND ISOLATED NODE OF INTERSECTION AREA BASED GEOCASTING PROTOCOL (IBGP) IN VANET

Sanjoy Das[1] and D.K Lobiyal[2]

[1,2]School of Computer and Systems Sciences, Jawaharlal Nehru University, New Delhi, India
[1]sdas.jnu@gmail.com , [2]lobiyal@gmail.com

## ABSTRACT

Geocasting is a special variant of multicasting, where data packet or message is transmitted to a predefined geographical location i.e., known as geocast region. The applications of geocasting in VANET are to disseminate information like, collision warning, advertising, alerts message, etc. In this paper, we have proposed a model for highway scenario where the highway is divided into number of cells. The intersection area between two successive cells is computed to find the number of common nodes. Therefore, probabilistic analysis of the nodes present and void occurrence in the intersection area is carried out. Further, we have defined different forwarding zones to restrict the number of participated nodes for data delivery. Number of nodes present and void occurrence in the different forwarding zones have also been analysed based on various node density in the network to determine the successful delivery of data. Our analytical results show that in a densely populated network, data can be transmitted with low radio transmission range. In a densely populated network smaller forwarding zones will be selected for data delivery.

.

## 1. INTRODUCTION

VANET is a special class of Mobile Ad hoc Network (MANET).A Mobile Ad hoc network is a dynamically reconfigurable wireless network with no fixed infrastructure. Every node in this network behaves like a router to relay a message from one node to another. In MANET; nodes are laptops, PDAs, palmtops, and other mobile devices whereas in VANET [1] nodes are vehicles. In addition, the other characteristics which differentiate VANET from MANET are mobility of nodes; structure of the geographical areas, delay constraint, privacy, etc. The node movement depends on the structure of road or structure of city or terrain etc. While delivering message from source to destination node, if destination node is not within the transmission range of source node then the source node sends the message to the destination node with the help of intermediate nodes. The ad hoc network is multi-hop in nature and message delivery depends on the connectivity among the intermediate nodes. The message delivery from one location to another location is done with the help of intermediate nodes. The aim of Geocast routing protocols is to deliver a message from one location (i.e. sender) to a predefined location known as Geocast region with optimal number of nodes and time period. It is desirable that protocols should maintain the low end-to-end delay and high success ratio in delivering message.

The rest of paper is organized as follows. Section 2 presents work related to the geocast protocols. In section 3 overview of our proposed mathematical model is presented. In section 4 through analysis of nodes presence and void occurrences has been analyzed in the intersection area. In section 5computation of different forwarding zones with node presence and void occurrence are discussed. Finally, the work presented in this paper is concluded in section 6.

## 2. RELATED WORK

Extensive works have been carried out by researchers, academicians and industries for successfully dissemination of messages from source to destination. There are several projects [2], [3], [4], [5] on VANET i.e. CarTalk, FleetNet–Internet on the Road, NoW (Network on Wheel)] are going on for its deployment in the real world. The main focus of all these projects is to provide safety, and timely dissemination of message from one location to another location. One of the message delivery protocols proposed for VANET tries to deliver a message to a geographic region rather than to a node called geocast routing. Many protocols have been developed for geocast routing such as LAR [6], LBM [7], GeoTORA [8],a modified TORA, GeoGRID [9], a modified GRID, GAMER [10], GRUV [11], etc. A Voronoi diagram based geocast protocol has been proposed in [12]. A comprehensive survey of geocasting protocol is presented in [13]. In [6, 7, 10, 11] authors used optimized flooding techniques for data delivery from source node to geocast region. In all these protocols, a forwarding zone is defined. The forwarding zone is a smaller area which includes source node, intermediate nodes and geocast region. The advantage of defining forwarding zone is that it helps in achieving optimized flooding and reduced routing overheads. Further, it also optimizes the network area. The data from source node is routed to geocast region with the help of intermediate nodes present within the forwarding zone. In [7] author's uses two types of forwarding zone LBM-box and LBM-step. The LBM-box is the smallest rectangle that includes source node and geocast region. The LBM-step is adaptive in nature it does not define the forwarding zone explicitly. The intermediate node which is closer to the smallest circle centered at the geometrical centre of the geocast region forwards the message. In [15,16] authors have divided the highway into cells. A probabilistic analytical model is proposed to examine the presence of neighbour nodes and isolated nodes on the highway. Analysis of the model shows that for the better network connectivity in multi hop VANET each cell should have at least one node. The numerical results show how connectivity of the network depends on the transmission range, network density and total network area.In [17] different forwarding zones are defined in the intersection area between two successive cells in a highway scenario. The result shows that in a densely populated network, data can be transmitted with low radio transmission range and vice versa for sparsely populated network. Further, selection of forwarding zone depends on the node density in the network.In [10, 11] authors have used different forwarding zones that are i) CONE ii) CORRIDOR and iii) FLOOD.  The GAMER provides mesh of paths between source node and geocast region. The mesh is created within the forwarding zone. As the traffic density in the network vary with time, both the protocols switches among different forwarding zones according to traffic density in the network. Once, a method fails to deliver message to the geocast region it automatically switches to other method. The main advantages of defining a forwarding zone are i) limited flooding in the network. ii) Reduced routing overheads iii) reduced overall network space. Each intermediate node receiving packet from source node will check its position first. If the receiving node falls within the forwarding zone, it only then forwards the packets to its next hop node, otherwise it discards the packet. None of these protocols considered the road structure since they have been primarily designed for MANET. In [14] authors have designed an analytical model for the performance analysis of Contention-based Geographic Forwarding (CGF) protocol with different forwarding areas. Further, they provide procedure for selecting different forwarding area for data transmission. The three forwarding areas named as Maximum Forwarding Area (MFA), Maximum Communication Area (MCA) and 60° Radian Area (DRA).The MFA is the overlapping area of two circular areas and it depends on the transmission radius, distance between the sender and the destination node.

The geocasting is the variant of multicasting, in geocasting source node sends message to a particular geographical area. We divided the geocast method in two phases. In the Phase-I, a source node sends a message to a node inside the geocast region. In the phase-II, the receiving node of the geocast region delivers the message to all the nodes in the geocast region. The node that moves to a geocast region automatically becomes the member of the geocast region. In Phase-II, the message delivery inside the geocast region is done by simple flooding techniques. Here, in this research we have confined our work only on the phase –I. In this work, we have carried out the probabilistic analysis of nodes and effect of various network parameters i.e., transmission range of node, node density on the performance of the network.

## 3. PROPOSED MODEL

We have considered multi-hop environment, because it's very rare that source and destination node fall within each other's transmission range. In case there is no direct connectivity between source and destination node, to route the message, intermediate nodes plays a vital role. The intermediate nodes act as relay or next forwarding node. We have considered highway scenario to deliver message from source to geocast region shown in Fig.1.

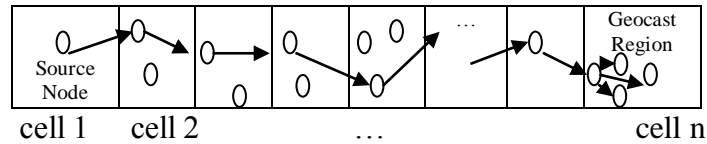

cell 1    cell 2    …    cell n

Figure1. Simple Scenario of Geocast on Highway [17]

| Symbols | Description |
|---|---|
| N | Total number of nodes in network |
| A | Area of network |
| R | Radio Transmission Range |
| $A_1$ and $A_2$ | Area within the intersection area |
| S | Source Node |
| $A_{int}$ | Intersection area between two successive cell |
| I | Intermediate Node |
| $\alpha_1, \alpha_2, \alpha_3$ | Radian Forwarding Zones |
| $\lambda$ | Node density per unit area |

Table 1. Symbol notation [17]

In our proposed model we have considered a highway scenario. It is divided in to *n* number cells of rectangular size. The length and width of the rectangular cell is denoted by *L*, and *W*, respectively. We assume that each cell is fully covered by a circular region of radius R. According to our assumption two successive cells share some common area that is denoted as intersection area. In Table 1. We have listed different symbols used in our analysis.

### 3.1. Computation of Intersection Area

In the Fig.1, we have shown a highway model for geocasting. We assume that at the center of each cell one node is present. The connectivity between center nodes depends on the nodes present within intersection area. Here, we compute the intersection area between two successive cells according to Fig.2.

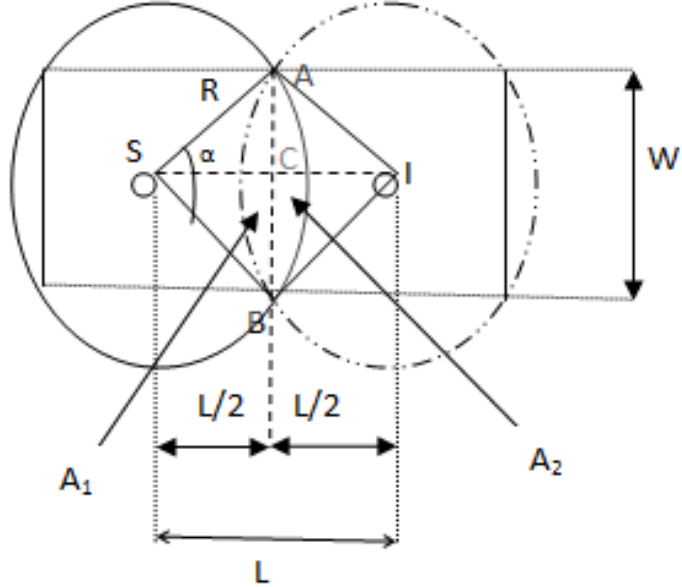

Figure 2. Shows the intersection area between two cells [17].

The Intersection area is denoted by $A_{int}$. Where, $A_{int}= A_1+A_2= 2 A_1$, because the area of $A_1$ and $A_2$ are equal. The area of $A_{int}$ computed as follows

The area of sector ABI=Area of sector SAB=$\frac{\alpha}{360} \times \pi R^2$ . (1)

The area of triangle SAB and ABI can be calculated as follows. For $\Delta$SAB, SA=SB=R and AB=W and for $\Delta$ABI, AI=BI=R and AB=W. According to Heron's area of a triangle it is calculated as

The semi-perimeter of $\Delta$SAB= S´= (SA+SB+AB)/2 = (2R+W)/2. (2)

$$Area^2 = S'(S' - SA)(S' - SB)(S' - AB)$$

$$Area^2 = \frac{(2R+W)}{2} \times \left(\frac{(2R+W)}{2} - R\right) \times \left(\frac{(2R+W)}{2} - R\right) \times \left(\frac{(2R+W)}{2} - W\right)$$

$$Area = \sqrt{\frac{(2R+W)}{2} \times \left(\frac{(2R+W)}{2} - R\right) \times \left(\frac{(2R+W)}{2} - R\right) \times \left(\frac{(2R+W)}{2} - W\right)}$$

$$= \frac{W}{4} \times \sqrt{4R^2 - W^2} .$$ (3)

Now, area of $A_1 = A_2 = \left(\frac{\alpha}{360} \times \pi R^2\right) - \left(\frac{W}{4} \times \sqrt{4R^2 - W^2}\right).$ (4)

$$A_{int} = A_1 + A_2 = 2 \times \left[\left(\frac{\alpha}{360} \times \pi R^2\right) - \left(\frac{W}{4} \times \sqrt{4R^2 - W^2}\right)\right]. \tag{5}$$

Where, $\tan\left(\frac{\alpha}{2}\right) = \frac{W}{L}$ So, $\alpha = 2 \times \tan^{-1}\left(\frac{W}{L}\right)$. Now compute the value of $A_{int}$ by putting the value of $\alpha$ in eq-(5).

$$A_{int} = 2 \times \left[\left(\frac{2 \times \tan^{-1}\left(\frac{W}{L}\right)}{360} \times \pi R^2\right) - \left(\frac{W}{4} \times \sqrt{4R^2 - W^2}\right)\right]$$

$$= \left(R^2 \times 2 \times \tan^{-1}\left(\frac{W}{L}\right)\right) - \left(\frac{W}{2}\sqrt{4R^2 - W^2}\right). \tag{6}$$

## 4. ANALYSIS OF PRESENCE OF NODES IN THE INTERSECTION AREA

We have considered a multi hop network scenario. If source and destination nodes are not in their direct communication range, intermediate nodes act as a relay node to deliver data from source to destination node. According to our model, we have chosen the intermediate node from the intersection area of two successive cells. The availability of nodes in the intersection area depends on various network parameters such as node density, radio transmission range, size of the cells, area of the network etc. The nodes in the network are distributed according to 2-D poisson point process. The probability of non- availability of nodes (void) in the intersection area can be calculated as follows. The probability of $m$ nodes present within an area A with average node density $\lambda$ is calculated as

$$p(m) = \frac{(\lambda A)^m \times e^{-\lambda A}}{m!}. \tag{7}$$

The probability of void in the area $A_{int}$ can be calculated as

$$p\{A_{int}(void)\}$$
$$= \frac{\left(\lambda \times \left(R^2 \times 2 \times \tan^{-1}\left(\frac{W}{L}\right)\right) - \left(\frac{W}{2}\sqrt{4R^2 - W^2}\right)\right)^0 \times e^{-\lambda \times \left(\left(R^2 \times 2 \times \tan^{-1}\left(\frac{W}{L}\right)\right) - \left(\frac{W}{2}\sqrt{4R^2 - W^2}\right)\right)}}{0!}$$
$$= e^{-\lambda \times \left(\left(R^2 \times 2 \times \tan^{-1}\left(\frac{W}{L}\right)\right) - \left(\frac{W}{2}\sqrt{4R^2 - W^2}\right)\right)}. \tag{8}$$

### 4.1. Numerical Results

In Fig.3, we have presented probability of void in the intersection area. The data transmission cannot be possible if no node is present in the intersection area. We have considered different number of nodes 50, 75,100, 150 and 200 with transmission range varying from 200 m to 500 m. We observed that, initially for transmission range of 200 m, as the number of nodes increases from 50 to 200, the probability of occurring void in the intersection area decreases from 0.2570to 0.0044. For transmission range 250 m the probability of occurring void in the intersection area becomes 0 for 150 and 200 nodes. For the simplicity we have shown the above result only for up to 200 nodes. The overall behaviour we observed that as the number of nodes increases probability of occurrence void is almost zero depends on the transmission range.

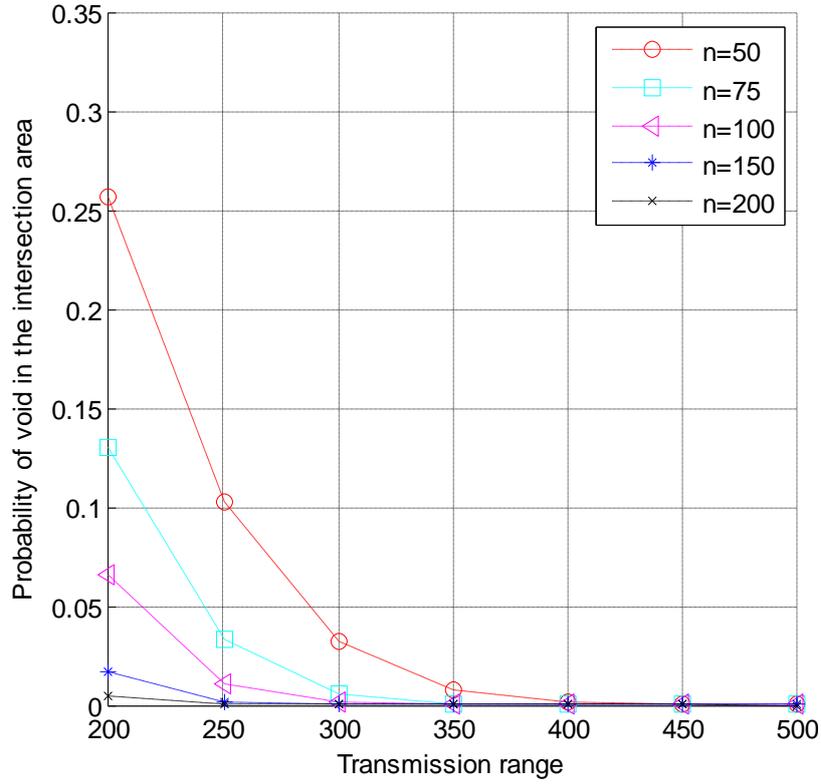

Figure 3. Probability of void in the intersection area with variable number of nodes.

| No of nodes | Transmission Range (m) | | | | | | |
|---|---|---|---|---|---|---|---|
| | 200 | 250 | 300 | 350 | 400 | 450 | 500 |
| 50 | 0.2570 | 0.1027 | 0.0325 | 0.0082 | 0.0016 | 0.0003 | 0.0000 |
| 75 | 0.1303 | 0.0329 | 0.0059 | 0.0007 | 0.0001 | 0.0000 | 0.0000 |
| 100 | 0.0661 | 0.0105 | 0.0011 | 0.0001 | 0.0000 | 0.0000 | 0.0000 |
| 150 | 0.0170 | 0.0011 | 0.0000 | 0.0000 | 0.0000 | 0.0000 | 0.0000 |
| 200 | 0.0044 | 0.0001 | 0.0000 | 0.0000 | 0.0000 | 0.0000 | 0.0000 |

Table 2. Probability of occurring void in the intersection region

In the above we have shown the probabilistic analysis of occurrence of void in intersection area. Now, we will analyze the presence of node in the intersection area by varying different network parameters. Therefore, the probability of m nodes present in $A_{int}$ area can be calculated as

$$p\{A_{int}(m)\} = \frac{\left(\lambda \times \left(R^2 \times 2 \times \tan^{-1}\left(\frac{W}{L}\right)\right) - \left(\frac{W}{2}\sqrt{4R^2 - W^2}\right)\right)^m \times e^{-\lambda \times \left(\left(R^2 \times 2 \times \tan^{-1}\left(\frac{W}{L}\right)\right) - \left(\frac{W}{2}\sqrt{4R^2 - W^2}\right)\right)}}{m!} \qquad (9)$$

Where, λ is the expected number of nodes within a unit area. In Fig.4, we shows probability of nodes present in the intersection area. We have considered different values of λ (i.e., 0.00004, 0.00005, and 0.00006) with fixed transmission range 250 m. We have presented the results only for 50 nodes for clarity of graphical representation. The probability of getting one node in the intersection area is 1, after 37 and 44 nodes when node density is 0.0004 and 0.0005. Further, for node density 0.0006, the probability of getting one node after 50 nodes onwards is one.

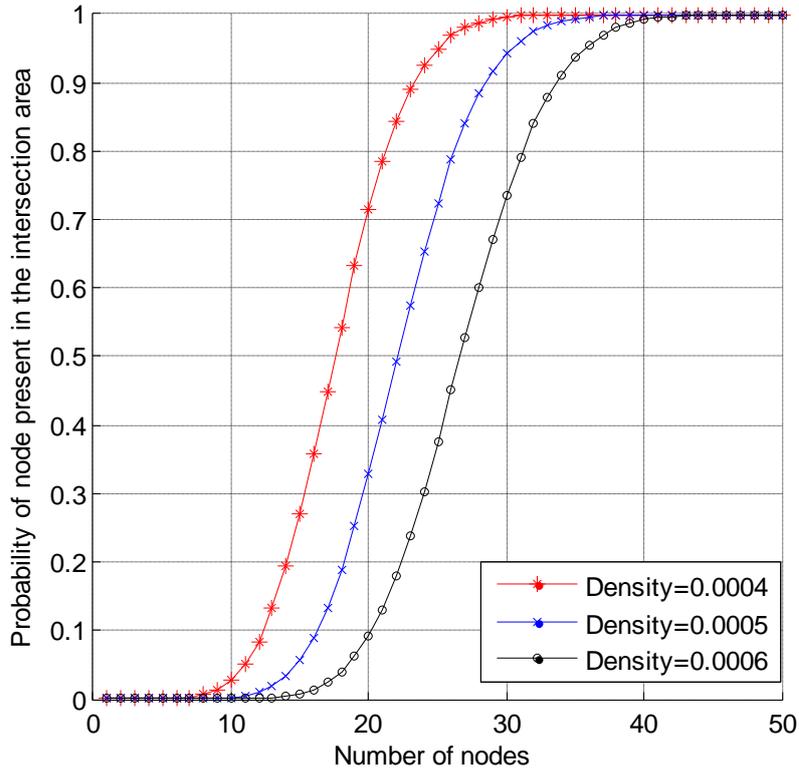

Figure 4. Probability of node present in the intersection area with variable node density.

As it is obvious and also observe from the above analysis that for the better connectivity of the network, intersection area should have at least one node present. In node density 0.0004 provides better connectivity in the network, because after 37 nodes there is always one node present in the intersection area. The higher the density of nodes, higher is the chance of getting one node in the intersection area. Therefore, higher density of nodes provides the better connectivity in the network. In Table 3. We have present the probability of node existing in intersection area with varying node density.

| Density (λ) | No of nodes | | | | | | |
|---|---|---|---|---|---|---|---|
| | 10 | 15 | 20 | 30 | 35 | 40 | 50 |
| 0.0004 | 0.0274 | 0.2706 | 0.7140 | 0.9961 | 0.9999 | 1.0000 | 1.0000 |
| 0.0005 | 0.0023 | 0.0572 | 0.3278 | 0.9424 | 0.9938 | 0.9996 | 1.0000 |
| 0.0006 | 0.0001 | 0.0076 | 0.0917 | 0.7357 | 0.9367 | 0.9914 | 1.0000 |

Table 3. Probability of node presence with various node densities in the intersection region.

## 5. FORWARDING ZONES OF IBGP

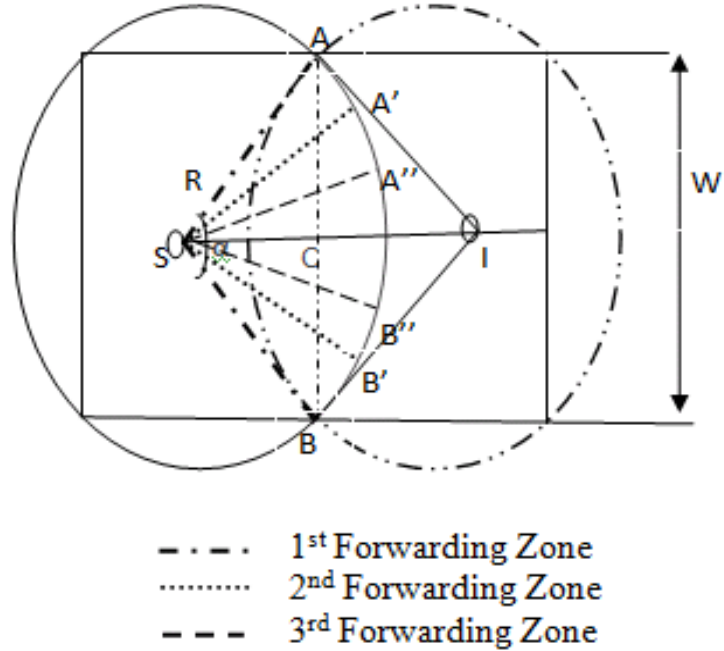

— · — · 1st Forwarding Zone
·········· 2nd Forwarding Zone
— — — 3rd Forwarding Zone

Figure 5. Forwarding zones of IBGP.

In our above discussion, we have given analysis of presence and absence of nodes in the intersection area for varying node density, and radio transmission ranges. Now, we have used different forwarding zones based on radian area according to [14]. The advantage of forwarding zone, it reduces the network overheads and network space for data delivery since the nodes that fall outside the forwarding zone do not participate in message delivery. In [14], a fixed radian area considered is of $60^0$. In our model we have chosen various value of radian area by varying the value of α shown in Fig.5. The value of radian area is changes with varying value of W and L.

### 5.1. Computation of Forwarding Zones

The radian area of angle α, i.e. the area of sector $SAB = \frac{\alpha}{360} \times \pi R^2 = \frac{2 \times \tan^{-1}\left(\frac{W}{L}\right)}{360} \times \pi R^2 = R^2 \times \tan^{-1}\left(\frac{W}{L}\right)$. Now we have defined three different radian forwarding zones according to Eq-(10).

$$RadianForwardingzone = \begin{cases} \alpha_1 \; ; \text{where } SAB = \tan^{-1}\left(\frac{W}{L}\right) \times R^2, AC = \frac{W}{2} \text{ and } SC = L/2 \\ \alpha_2 \; ; \text{Where } SA'B' = \tan^{-1}\left(\frac{W}{2L}\right) \times R^2, AC = \frac{W}{4} \text{ and } SC = L/2 \\ \alpha_3 \; ; \text{Where } SA''B'' = \tan^{-1}\left(\frac{W}{4L}\right) \times R^2, AC = \frac{W}{8} \text{ and } SC = L/2 \end{cases} \quad (10)$$

Probabilistics analysis of void occurance in the different forwarding area can be calculated as

$$p\{\text{RadianForwardingzone } \alpha_1 \text{ (void)}\} = e^{-\lambda \times \left(\tan^{-1}\left(\frac{W}{L}\right) \times R^2\right)}. \quad (11)$$

$$p\{\text{RadianForwardingzone } \alpha_2 \text{ (void)}\} = e^{-\lambda \times \left(\tan^{-1}\left(\frac{W}{2L}\right) \times R^2\right)}. \quad (12)$$

$$p\{\text{RadianForwardingzone } \alpha_3 \text{ (void)}\} = e^{-\lambda \times \left(\tan^{-1}\left(\frac{W}{4L}\right) \times R^2\right)}. \quad (13)$$

### 5.2. Numerical Results

In Fig.6 shows the occurrence of void in different forwarding zones. We have fixed the number of nodes as 500 and vary the transmission range from 250 m to 500 m. The probability of occurrence of void in different forwarding zones is clearly shown in Fig 6(a), Fig 6(b) and Fig 6(c). In first forwarding zone the probability of occurrence of void is 0.5099 x10$^{-7}$ for transmission range of 250 m. In 2$^{nd}$ and 3$^{rd}$ forwarding area probability of occurrence of void is 0.4865 x10$^{-22}$ and 0.4926 x10$^{-22}$, respectively for transmission range 250 m. The probability of occurrence of void is zero for 1$^{st}$ forwarding zone for transmission range 320 m and for 2$^{nd}$ and 3$^{rd}$ zone is zero after transmission range 280 m. In 1$^{st}$ forwarding area to provide better connetcivity in the network data should be transmitted using 320 m transmission range or above and for the other two zones it should be higher than 280 m.

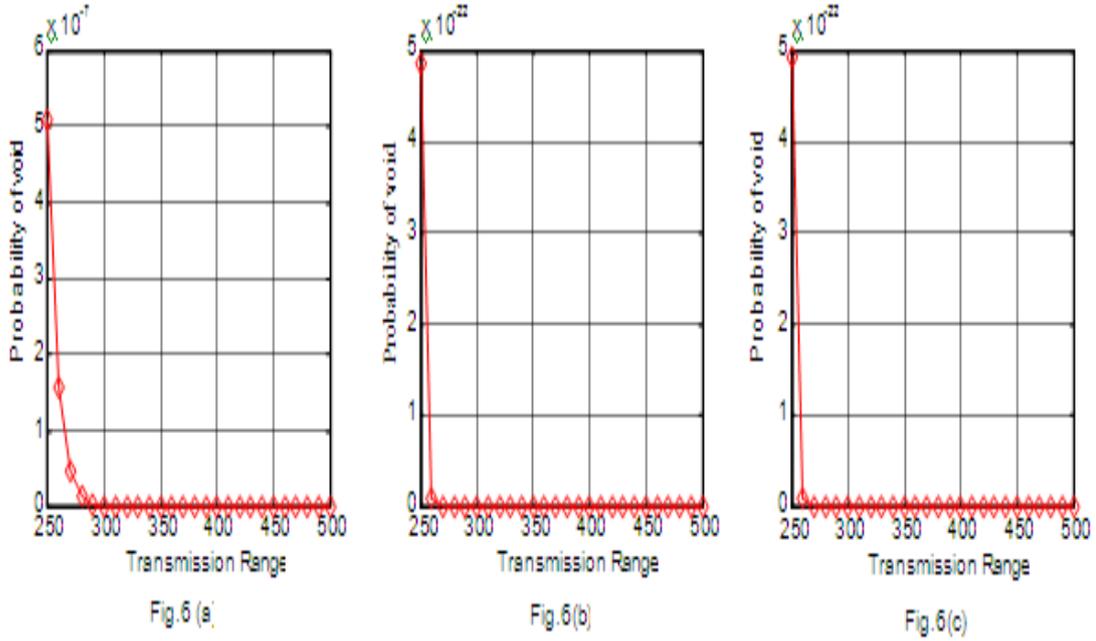

Figure 6. Probability of void in different Forwarding Zones.

Probabilistics analysis of presence of nodes in different forwarding zones can be calculated as

$$p\{\text{RadianForwardingzone}\alpha_1 \text{ (m)}\} = \frac{\left(\lambda R^2 \times tan^{-1}\left(\frac{W}{L}\right)\right)^m \times e^{-\lambda\left(R^2 \times tan^{-1}\left(\frac{W}{L}\right)\right)}}{m!}. \quad (14)$$

$$p\{RadianForwardingzone\alpha_2 (m)\} = \frac{\left(\lambda R^2 \times tan^{-1}\left(\frac{W}{2L}\right)\right)^m \times e^{-\lambda\left(R^2 \times tan^{-1}\left(\frac{W}{2L}\right)\right)}}{m!}. \quad (15)$$

$$p\{RadianForwardingzone\alpha_3 (m)\} = \frac{\left(\lambda R^2 \times tan^{-1}\left(\frac{W}{4L}\right)\right)^m \times e^{-\lambda\left(R^2 \times tan^{-1}\left(\frac{W}{4L}\right)\right)}}{m!}. \quad (16)$$

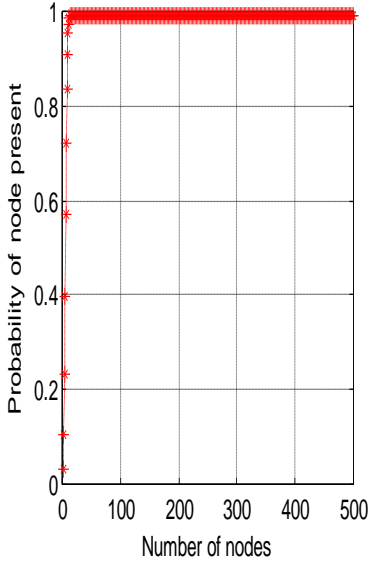
Fig.7(a)

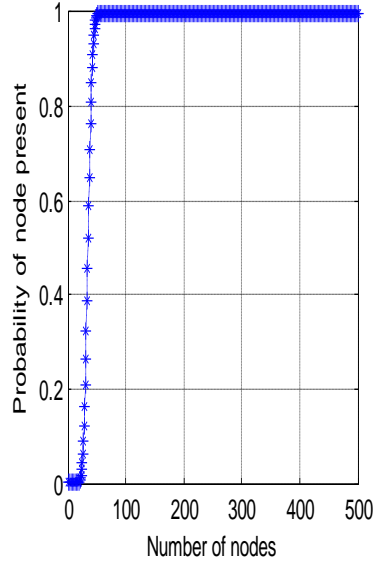
Fig.7(b)

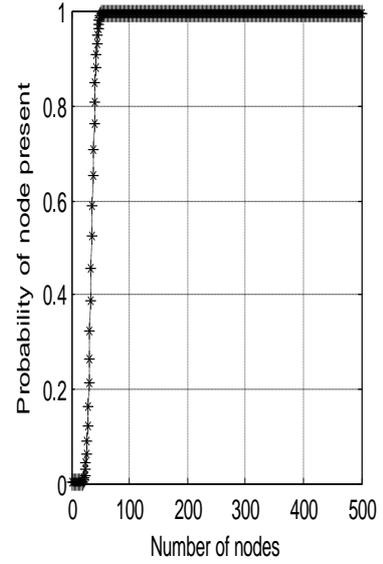
Fig.7(c)

Figure 7. Probability of node presence indifferent Forwarding zones.

In Fig . 7 shows the probability of presence of nodes in different Forwarding zones for the fixed transmission range of 150 m. We have presented the result only for up to 500 nodes.In Fig.7(a), after 17 nodes in the first forwarding zone nodes are always present and its probability is 0.9946. In Fig.7(b) and Fig.7(c) till 14 nodes there is no node present but after 61 nodes, there are always nodes present in $2^{nd}$ and $3^{rd}$ forwarding zone.Therefore, we can observed that in low density network $1^{st}$ forwarding zone should be considered and when the number of nodes are above 60 nodes, $2^{nd}$ and $3^{rd}$ forwarding zone should be considered. In Table 4. We have presented the data for probability of nodes existing in different forwarding zones.

| Forwarding Zones | No. of Nodes | | | | | | | | |
|---|---|---|---|---|---|---|---|---|---|
| | 5 | 10 | 20 | 30 | 40 | 50 | 100 | 150 | 200 |
| I | 0.5727 | 0.9765 | 0.9946 | 0.9946 | 0.9946 | 0.9946 | 0.9946 | 0.9946 | 0.9946 |
| II | 0.0000 | 0.0000 | 0.0037 | 0.2107 | 0.8096 | 0.9923 | 1.0000 | 1.0000 | 1.0000 |
| III | 0.0000 | 0.0000 | 0.0037 | 0.2111 | 0.8101 | 0.9923 | 1.0000 | 1.0000 | 1.0000 |

Table 4. Probability of node presence in different Forwarding Zones.

## 6. CONCLUSION

In this paper we have analyzed the impact of node density, transmission range, network area on the connectivity of the network for a highway scenario. In this work we have computed the probability of presence of common nodes in the intersection area of successive cells in a linear route. The higher the density of nodes, higher is the chance of getting one node in the intersection area. Therefore, higher density of nodes provides the better connectivity in the network. We have also investigated the impact of three different forwarding zones and their selection is depends on node density. In a densely populated network smaller forwarding zones will be selected for data delivery. From the analysis of simulation results for higher number of nodes and transmission range, the probability of void occurrence in the intersection area is low. But, for better network connectivity and uninterrupted data transmission intersection area should have at least one node. Further, it also suggests that if the number of nodes is less, the probability of void occurrence is high.

## REFERENCES

[1] Moustafa, Hassnaa., Zhang, Yan. (2009) Vehicular networks: Techniques, Standards, and Applications, CRC Press, Taylor & Francis Group, Boca Raton London, New York.

[2] The NoW: Network on wheels Project. http://www.network-on-wheels.de/about.html

[3] CarTalk, http://www.cartalk2000.net/

[4] FleetNet–Internet on the Road, http://www.fleetnet.de/

[5] http://vanet.info/projects

[6] Ko, Y.B., &Vaidya, N. (1998) "Location-aided routing (LAR) in mobile ad hoc networks", In Proceedings of the ACM/IEEE International Conference on Mobile Computing and Networking (MOBICOM'98), pp. 66–75.

[7] Ko, Y.B., &Vaidya, N. (1999) "Geocasting in Mobile Ad-hoc Networks: Location-Based Multicast Algorithms", In 2nd IEEE Workshop on Mobile Computing Systems and Applications, New Orleans, Louisiana.

[8] Ko, Y.B., &Vaidya, N. (2000) "GeoTORA: A Protocol for Geocasting in Mobile Ad Hoc Networks", In: IEEE International Conference on Network Protocols, Osaka, Japan, pp.240-250.

[9] Liao, W.H., Tseng, Y.C., & Sheu, J.P. (2000) "GeoGRID: A Geocasting Protocol for Mobile Ad Hoc Networks Based on GRID", *Journal of Internet Tech.* Vol.1 (2), pp. 23–32.

[10] Camp,T.,& Liu,Y.(2003)"An adaptive mesh-based protocol for geocast routing", *Journal of Parallel and Distributed Computing: Special Issue on Routing in Mobile and Wireless Ad Hoc Networks*, Vol. 62(2), pp.196-213, (2003).

[11] Zhang, Guoqing.,Chen, Wu., Xu,Zhong., Liang, Hong.,& DejunMu,LiGao (2009) "Geocast Routing in Urban Vehicular Ad Hoc Networks", In Lee, R., Hu, G.,Miao , H., (eds.) *Computer and Information Science 2009*, Vol. 208, Springer-Verlag Berlin Heidelberg,pp. 23–31.

[12] Stojmenovic, I., Ruhil,A. P., & Lobiyal, D. K. "Voronoi diagram and convex hull based Geocasting and routing in wireless networks", In Proc. of the 8th IEEE Symposium on Computers and Communications ISCC, Antalya, Turkey, pp. 51-56, (2003).

[13] Maihofer, C. (2004) "A Survey of Geocast Routing Protocols", *IEEE Communications Surveys &Tutorials* ,Vol.6(2), pp.32–42.

[14] Chen,D.,Deng, J., & Varshney, P. K. (2007) "Selection of a Forwarding Area for Contention-Based Geographic Forwarding in Wireless Multi-hop Networks", *IEEE Transactions on Vehicular Technology,* vol. 56, no. 5, pp. 3111-3122.

[15] Das, Sanjoy., & Lobiyal, D. K. (2011)"An Analytical Analysis of Neighbour and Isolated Node for Geocast routing in VANET",*International journal on Ad Hoc networking Systems ( IJANS),* Vol. 1, No. 2, 2011, pp. 39-50.

[16] Das, Sanjoy., & Lobiyal, D. K. (2011)"Analysis of Next Hop Selection for Geocasting in VANET", D.Nagamalai, E.Renault, M.Dhanushkodi (Eds.): CCSEIT-2011, CCIS Vol. 204, part 1, pp.326-335, 2011, Springer-Verlag Berlin Heidelberg. DOI: 10.1007/978-3-642-24043-0_33.

[17] Das,Sanjoy., & Lobiyal, D. K. (2012) "Intersection area based geocasting protocol (IBGP) for Vehicular Ad hoc Networks", N. Meghanathan et al. (Eds.): CCSIT 2012, Part III, LNICST 86, pp. 387–396, 2012. Institute for Computer Sciences, Social Informatics and Telecommunications Engineering (LNICST), Springer-Verlag Berlin Heidelberg.

**Authors**

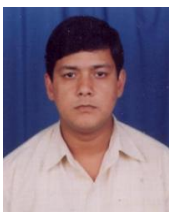

Sanjoy Das received his B. E. (Computer Science and Engineering) from G. B. Pant Engineering College, Pauri-Garhwal, U.K, India and M. Tech (Computer Sc. & Engg.) from Sam Higginbottom Institute of Agriculture, Technology and Sciences, Allahabad (U.P), India in 2001 and 2006, respectively. He is pursuing Ph.D in Computer Science as a full time research scholar in the School of Computer and Systems Sciences, Jawaharlal Nehru University, New Delhi, India. He had worked as an Assistant Professor in the Department of Computer Science and Engineering in G. B. Pant Engineering College, Uttarakhand Technical University, India and also in the department of Information Technology, School of Technology, Assam University (A Central University), Silchar, Assam, India. His current research interest includes Mobile Ad hoc Networks and Vehicular Ad hoc Networks.

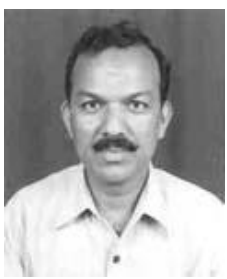

*Daya K. Lobiyal* Received his Bachelor of Technology in Computer Science from Lucknow University, India, in 1988 and his Master of Technology and Ph.D both in Computer Science from Jawaharlal Nehru University, New Delhi, India, 1991 and 1996, respectively. Presently, he is an Associate Professor in the School of Computer and Systems Sciences, Jawaharlal Nehru University, India. His areas of research interest are Mobile Ad hoc Networks, Vehicular Ad Hoc Networks, Wireless Sensor Network and Video on Demand.